# Volumetric based mass flow estimation on sugarcane harvesters

Muhammad K.A. Hamdan[a] · Diane T. Rover[a] · Matthew J. Darr[b] · John Just[b]

[a] Department of Electrical and Computer Engineering
Iowa State University, Ames, Iowa, USA
[b] Department of Agriculture and Biogas Engineering
Iowa State University, Ames, Iowa, USA

## Abstract

Yield monitors on harvesters are a key component of precision agriculture. Mass flow estimation is the critical factor to measure, and having this allows for field productivity analysis, adjustments to machine efficiency, and cost minimization by ensuring trucks are filled maximally without exceeding weight limits. Several common technologies used on grain harvesters, including impact plate sensors, are accurate enough on combines to be valuable but suffer from issues such as drift. Sugarcane is composed of a mixture of billets and trash, which is a very dispersed material with much less consistency than grains. In this study, a 3d point cloud approach is used to estimate volume, from which a calibration factor is derived [density] to translate to mass. The system was proved in concept in a controlled environment using bamboo, achieving an $R^2$ of 97.4% when fitting average volume flow per test against average mass flow after correcting for bulk density changes with volume. The system was also tested on field data, which was collected from nearly 1700 wagon loads from the southern U.S. and Brazil over the course of 3 seasons in both green and burnt cane. Results indicated that the concept is very robust with good accuracy, having seasonal CVs for density values ranging from 6.9% to 16.2%. The camera concept proves relatively robust to environmental conditions. The same approach could be used in sugar beets, potatoes or other sparse/non-flowing crops with highly varying material properties, where traditional mass flow sensors do not work.

Corresponding author
Iowa State University, Ames IA, USA
✉ E-mail: mhamdan@iastate.edu

# 1 Introduction

Sugarcane is an established source of sugar and it is one of the feed stocks for efficient biofuel production (Somerville et al. 2010). Originally grown in Southeast Asia, sugarcane is now grown in tropical and subtropical countries around the world, including Brazil and a portion of the U.S. As a ratoon crop, it can be grown for 5 to 7 years and harvested annually without replanting (Magness, Markle, and Compton 1971). Sugarcane can even be grown for longer periods if not damaged during harvesting such that it continues to regrow. The crop grows year-round, regenerating after cutting, and is usually re-harvested at the same time period in the following year. With this high throughput production, a means for yield estimation is needed to support the operations and planning. A yield monitor is a device consisting of a coupled set of sensors, which measure geographic and mass flow information. during harvesting. This information can in-turn be translated to crop yield spatially within a field or over time

As human population continues to increase, pressure on the agricultural supply chain will increase. The same land must produce more food, while global competition pressures producers and suppliers to increase efficiencies in all areas of operations. Adoption of current precision agriculture tools and technologies has led to higher yields, reduced environmental impact, reduced costs, and improvements in sugarcane quality (Silva, de Moraes, and Molin 2011) . Yield monitors have been a core component of precision agriculture. Some of the earliest sugarcane yield monitors were developed in USA by Ag leader technology (Ag Leader 1992).

As an essential component of a yield monitor, a mass flow sensor provides feedback related to the amount of product at each point in the field where the harvester operates. This enables farmers to assess the performance in and between fields, and supports quantitative decisions such as when to replant or how to adjust inputs such as fertilizers. Additionally, mass flow sensors support real-time control of machine productivity by monitoring machine work against quantity harvested, or help in optimizing logistics such as filling semi-tractor trailers maximally without exceeding allowed limits.

There have been several techniques attempted over the last 20 years while trying to measure sugarcane yield on a machine during harvesting. Direct methods to yield estimation include using an impact or deflection plate similar to what is ubiquitously used in grain crops, or a weigh plate in the elevator floor that triggers a reading as each slat passes by. Indirect methods include measuring and calibrating to machine functions (e.g. power consumption), or the use of non-contact sensor. Realizations of systems following the direct and indirect methods have been attempted with varying results by several researchers. (G. J. Cox 2002) used chopper and feed roller hydraulic pressure (indirect mass measurement) along with motor speed to relate power exerted by the motors and volume of material through rollers to mass flow via a regressed calibration. While simple and inexpensive since it requires very little change to the machine, wear on the machine and different crop conditions may all shift the response. (Mailander et al. 2010) designed a monitoring system that primarily uses a scale to measure the yield of sugarcane. The system resulted in an average percent error of 11%.(Graeme Cox et al. 1996), (G Cox, Harris, and Pax 1997), (Pagnano and Magalhaes 2001), (Molin and Menegatti 2004), (Cerri and Magalhães 2005), and (Magalhães and Cerri 2007) tried a version of a weigh plate with load cells in the floor of the elevator. Adjustments for pitch of the elevator were made with an accelerometer, and readings triggered as slats passed over the plate. It is a direct mass measurement but it requires significant changes to the machine, is costly and complex, and is susceptible to mechanical noise. Additionally, load cell drift due the moving parts of the harvester and can suffer from buildup of material jamming the plate.

Other methods that have been investigated to estimate sugarcane yield include force impact. (Wendte, Skotnikov, and Thomas 2001) described a yield monitor system that uses force to estimate yield of cane through utilizing a deflection plate sensor and a control monitor. The deflection plate is placed near to the top end of the elevator after the billets exit to fall into the wagon. The measured force and elevator speed are calibrated to predict mass flow. Another approach to estimate yield by measuring the stem-bending force was tested by (Mathanker et al. 2015). Load cells were placed between two parallel pipes making a push bar that was installed between crop dividers. The system showed an $R^2$ of 92% between forces on the push bar and yield when harvesting napiergrass, but the system could not withstand the large impact forces encountered during the harvest of sugarcane.

An under-elevator optical sensor array was test by (Randy R Price, Larsen, and Peters 2007). They measure the amount of time the sensor array was covered with material and then calibrate the calculated duty-cycle to mass flow. This is an indirect volumetric method which is relatively simple and inexpensive. However, it depends on lighting and a reliable stacking of material since it attempts to quantify volume using only two dimensions. Additionally, the calibration is highly affected by changes in material density. Further, (R R Price et al. 2011) developed a fiber optic yield monitoring system that measures the volume of sugarcane to estimate yield. Under dry field conditions the system resulted in an average error of 7.5%; however, the system performance degraded under wet and muddy conditions, where about 75% of sensor readings were lost. (Randy R Price, Johnson, and Viator 2017) also developed an alternative optical yield monitor system that uses two laser distance sensors mounted above the loading elevator to measure height and length of the billet piles per slat. Different methods were attempted to find the best relationship between volume and material weight. It was found that the cumulative billet pile length had the best relationship to harvested weight. An $R^2$ ranging from 93% to 97% was reported. The system is relatively simple and easy to install; however, the system still suffered from the piling up of debris and sugarcane leaves.

To our knowledge, there has not been research that used an indirect volumetric based approach with point clouds to estimate the yield of sugarcane; however, (Jadhav et al. 2014) developed a volumetric mass flow sensor that uses a LiDAR sensor to estimate the total mass of citrus. Their system was tested on an inclined and a horizontal conveyor like those used on mechanical harvester and debris removal systems. Results show that the system can estimate mass flow with an average error of 7% and a standard deviation of 7% for the incline conveyor, and an average error of 7% and a standard deviation of 5% for the horizontal conveyor. Several direct and indirect volume-based measurement systems were also explored by (Schmittmann, Oliver 2001) in a small trial for sugar beet and potato yield. They reported good results with a laser scanner (< 4% error) and mechanical fingers, but no further information could be found of a product developed from their work.

In this study, a volumetric approach is extensively explored to estimate sugarcane yield by generating a point cloud of the material on the elevator using a stereo camera. Stereo cameras create a full 3d point cloud of the material, unlike laser scanners which capture a 2d plane with each measurement and miss material in be-tween each measurement. The objective of the work is to create and extensively validate a robust and reliable yield monitor for sugarcane that can be extended to other similar crops without adding mechanical complexity.

# 2 Materials and Methods

## 2.1 Fundamentals of Operation (Theory)

The basic operation of the yield monitor proposed herein consists of measuring the volume of the flowing material (in this case sugarcane billets) on the machine elevator during harvesting, and converting the measured volume to mass via a predicted density value. This is done using a stereo camera instead of Lidar for cost effectiveness, robustness to dust and dirt, and a much larger measurement area which does not miss material.

For comparison, a Lidar line would measure an x-z cross section of the elevator (in the same plane as the slats) to obtain depth ("z") measurements at equally spaced horizontal ("x") distances across the width of the elevator. At a typical elevator speed of 2 m/s, a sensor sampling at 25Hz will only acquire a cross-sectional measurement of the material spacing of 8cm ($200\frac{cm}{s} \times \frac{1}{25}s$) along the direction that the material is conveyed on the elevator. Thus it would be necessary to interpolate (8cm to 20cm, where 20cm is the distance between nearby slats) in the "y" direction between x-z cross sectional measurements to create a surface from which to calculate volume. A stereo camera, on the other hand, produces a dense 3d point cloud over a large surface with each measurement and facilitates better resolution of the surface of the material (i.e. no need for interpolation). In this case the system operated at 7.5Hz which was enough for overlapping successive point clouds and ensuring no information loss.

While the stereo camera can have errors in estimation of volume, laboratory testing in this work showed highly consistent results with the stereo cameras as long as lighting was maintained above a critical level, which was easily achieved with supplementary LED lighting. Once a volume measurement is obtained, then density can be predicted and used to translate the volume measurement into mass. Grains are known to expand and contract with moisture, and change both the particle and bulk densities. In the case of highly non-uniform shape characteristics like sugarcane billets, which are long slender rods, bulk density has the potential to be more complex from the shape factor alone, along with particle density expected to additionally contribute to density complexity.

The for mass flow, up to a scaling constant depending on the time period to estimate over, is shown in Equation 1. The stereo camera estimates the volume within the region of interest (ROI), which is denoted as $V_c$ and has units of meters cubed per meter up to a scaling constant. Then this quantity is scaled by the distance the elevator moves ($\Delta t \times V_e$) in between the next volume estimate to find an incremental accumulated volume. A simplifying assumption inherent in this formulation is that the volume calculated ($V_c$) is spread evenly across the ROI, since the incremental accumulated volume $V_\Delta$ is directly proportional to the distance the elevator moves. The incremental volume is then converted to mass via a multiplier (density) as seen in Equation 1. Any error in density of the material can be seen to directly contribute to error in predictions of mass, and therefore yield as shown in Equation 2.

$$V_\Delta = \Delta t \times V_e \times V_c \; ; \; m_\Delta = V_\Delta \times \rho \qquad (1)$$

where:

$V_\Delta$ = incremental accumulated volume

$V_c$ = volume from stereo camera

$\Delta t$ = image capture time

$V_e$ = harvester elevator velocity level

$m_\Delta$ = incremental accumulated mass

$\rho$ = density (calibration factor)

$$Yield = \frac{\dot{m}}{w \times V_m} \tag{2}$$

where:

$\dot{m}$ = mass flow

$V_m$ = vehicle speed

w = row width

## 2.2 Laboratory Setup

Extensive proof of concept testing was conducted prior to field exposure. Experiments were designed to test the system as closely as possible to factors present during typical operation, while controlling for factors outside of system control such as particle density. To accomplish this, bamboo was used as a surrogate material to sugarcane since it has stable material properties long term (will not rot or dry out) while being similar in shape to sugarcane. Bamboo was conveyed into a sugarcane elevator at various flow rates with a stereo camera mounted on top of the elevator. Figure 1 shows the overall system setup, which includes a hydraulically driven elevator and conveyor, a logging system, a stereo camera system, and a scale. Initially, a weight measurement (ground truth) is taken using the scale, then the material is conveyed via the conveyor and elevator. As the material travels, the stereo camera produces a 3d point cloud of the material that is converted to volume via a matching algorithm. Lastly, the logging system records the volumetric data and stores it on a storage unit.

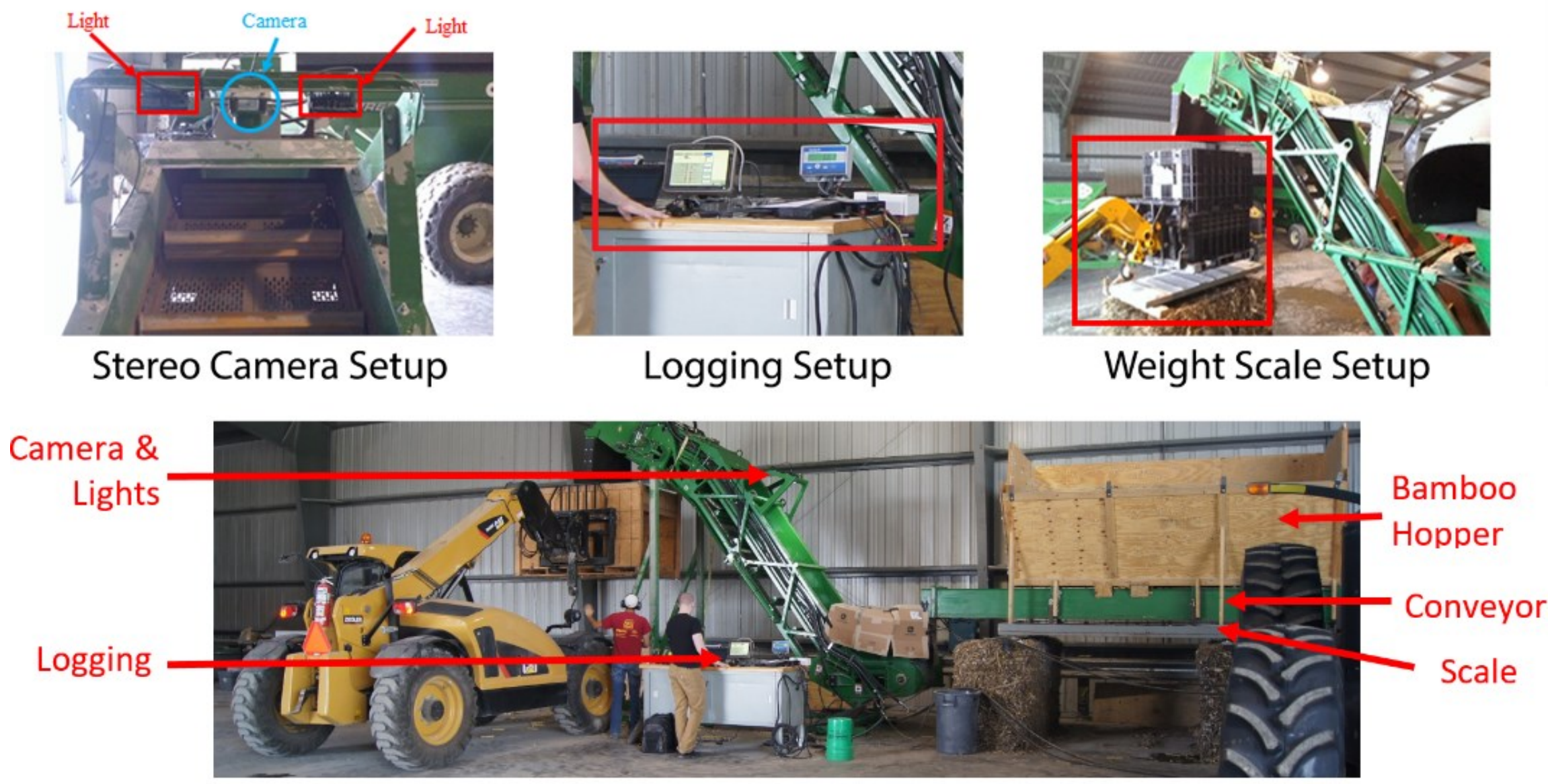

**Fig. 1** A picture of the overall system showing the laboratory setup for volume measurement testing.

The design of experiment (DOE) took into account various factors such as variable illumination, mass flows, and elevator speeds. These factors were projected to impact the response of the measurements and simulate environmental variables of operating conditions in a commercial application. Six lighting levels were used that span from ≈ 0.7k lux to ≈ 6.7k lux as measured by a light meter located on the elevator in front of the camera. The elevator speed was measured by an inductive proximity sensor on the drive sprocket and varied between 1.0 m/s to 2.2m/s. Runs length spanned between 20 seconds up to 120 seconds and the total mass per run ranged between 230kg up to 300 kg. A total of 239 runs were obtained, which include 8 empty runs (zero mass) with the elevator moving to be sure the system was unbiased.

### 2.3 Field Setup and Data Summary

The laboratory and field systems are identical except for the wireless transceivers used to collect ground truth from scales installed on the wagons as shown in the diagram in Figure 2. The wagons were typically six or nine metric ton capacities. An image processing unit (stereo camera + algorithm) was used to generate 3d point cloud of material and convert into volume, and a speed sensor was used to capture the elevator speed so to scale the volumetric data. The image processing unit was controlled via CAN bus and data was transferred to a dedicated stereo logger via an Ethernet link between the image processing unit and the stereo logger.

Field data presented in this work was collected under real operation conditions, and is summarized by crop type, season of harvest, and region in Table 1. Data was collected in the course of 3 consecutive years (2014 through 2016) and in 4 different regions (Brazil, Florida, Louisiana, and Texas) to ensure robustness of the system to various environmental factors that could influence bulk and particle densities of the material. Data is comprised of 1567 runs (over 3M frames), and is split between burnt and green cane. This was an important part of the DOE since burnt cane, in which the leaves and fibrous trash were burnt off, was projected to represent the high end for density, whereas green cane was expected to vary more depending on the amount of trash and ability of the primary extractor fan to remove trash.

**Table 1** Runs distribution of years and region of harvest

| | | Cane Harvest Year | | | |
|---|---|---|---|---|---|
| **Crop Type** | **Region** | **2014** | **2015** | **2016** | **Total/Region** |
| | **Louisiana** | 0 | 669 | | 669 |
| **Green** | **Brazil** | 166 | | | 166 |
| | **Florida** | | | 264 | 264 |
| | **Texas** | 66 | 79 | | 145 |
| **Burnt** | **Florida** | | | 323 | 323 |
| **Total per year** | | 232 | 748 | 587 | Total: 1567 |

Early on in testing (2014, and Texas 2015), ground speed and fan speed were used to induce mass flow changes as well as changes in material composition (and therefore density). While they are useful for obtaining more variation in the dataset, they are not globally consistent factors since the magnitude of the effect of ground speed on mass flow and fan speed on material composition depends on yield and trash levels, respectively.

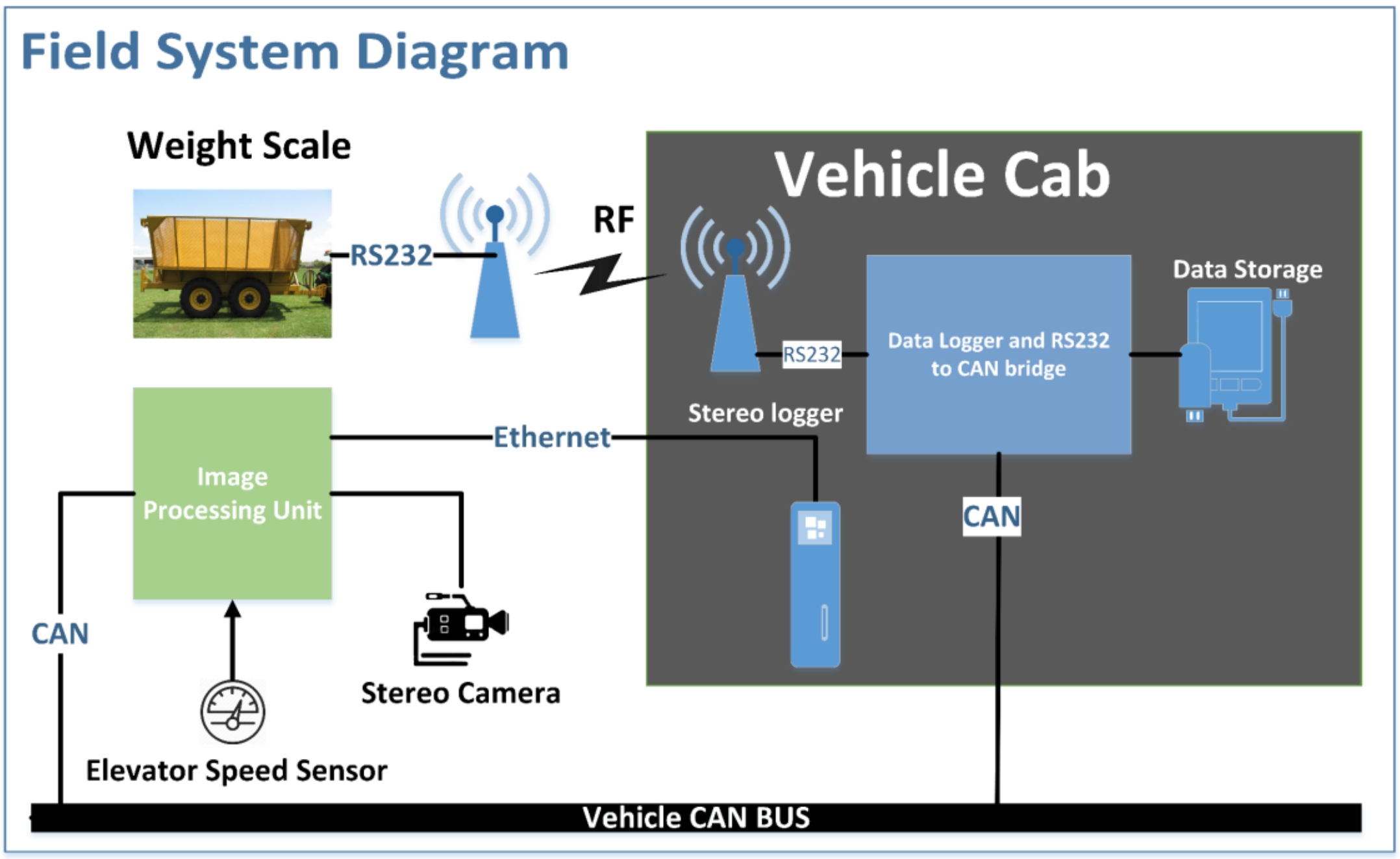

**Fig 2** Block diagram of the field system setup highlighting the components used for data generation and logging.

## 2.4 Mass Estimation

In this system we have instantaneous volumetric measurements and ground truth mass measurements per run (video). To estimate instantaneous mass we first need to estimate the underlying density [calibration factor]. Density is calculated as the total ground truth mass of the material divided by the accumulated volume (sum of $V_\Delta$ from Equation 1). Applying a square root transformation to the volume measurements prior to estimating the density was found to improve mass estimates through reducing data dispersion. Once density is estimated, it is then used to produce instantaneous mass measurements per frame as shown in Equation 3.

$$m_\Delta = \rho \times sqrt(V_c) \times V_e \times \Delta t \qquad (3)$$

In this formulation $\rho$ is the underlying estimated density and "*sqrt*" is a square root transformation that is applied to volume ($V_c$), scaled by elevator speed ($V_e$) and time between measurements ($\Delta$t).

# 3 Results and Analysis

## 3.1 Controlled Lab Testing

As per the design of experiments, lighting was one factor that was controlled during testing. It was found during laboratory testing that so long as lighting was kept at a certain minimum level (>2.7*k* lux), significant nonlinearities with the point cloud estimation were avoided. In low lighting, the stereo camera matching algorithm was found to produce a higher number of points for the point cloud overall, but the points tended to have a lower-than-expected elevation as measured from the plane of the elevator. Since the shutter speed had to be fast due to the quick movement of the slats (and therefore material), increasing the exposure time of the camera was not an option.

The data points circled in red in Figure 3.a were collected under light levels below 2.7k lux, hence a significant decreasing density trend is observed as the volume increases. Thus far, as long as certain minimum amount of lighting is provided, a decreasing density still persists but is not as extreme as shown by the data points circled in green in Figure 3.b. This remaining decreasing density trend was hypothesized due to a packing effect where the measured bulk density of the material decreases as the total volume of material increases. Work investigating the densities of disordered packing of slender cylinders supports this (Zhang 2006) and (Liu et al. 2018).

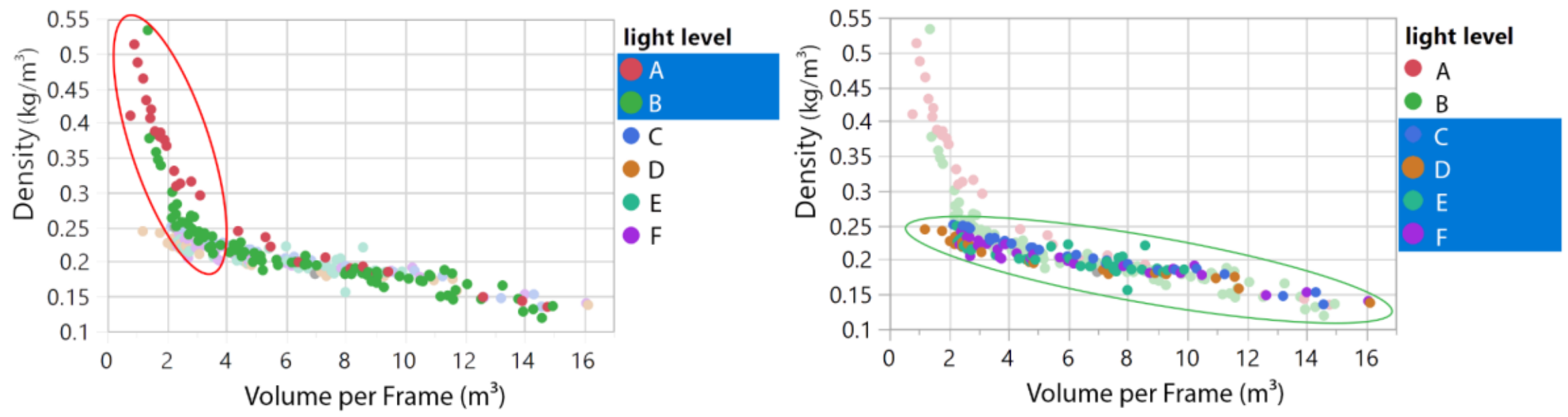

(a) Light levels < 2.7*K* lux

(b) Light levels > 2.7*K* lux

**Fig. 3** Left: Density dependence trends. Light levels (A & B): poor sensitivity of stereo camera at lower volume flows with light level less than 2.7 lux. Right: Light levels (C, D, E, & F): runs with adequate lighting show decreasing bulk density with volume likely due to disordered stacking of material – a real/physical phenomenon.

Note that the variation increases for volume flows larger than 12m$^3$ per frame since at that level bamboo begins to overflow the slats and fall back down the elevator, resulting in double counting and lower fidelity point cloud estimation. This can also be seen in the circled points in Figure 4.a. Bamboo is also especially slick when compared with sugarcane, which is less likely to experience this issue due to higher friction and generally not often reaching such high flow rates.

We consider evaluating the coefficient of variation (CV) of density values, since density is the calibration factor that is used to predict mass. The CV of the density values was 12% and transforming the volume shaved 2%, bringing it from 12% to 10%, with the histogram comparison shown in Figure 4.b. It is important to minimize the coefficient of variation of density values since the lower the value of the coefficient of variation, the more precise the estimate. Since environmental factors like lighting are significantly more variable in the field application, and material composition is far more inconsistent with leaves, billets, and moisture content, a CV of 10% was deemed to represent a good lower-bound expectation for product performance.

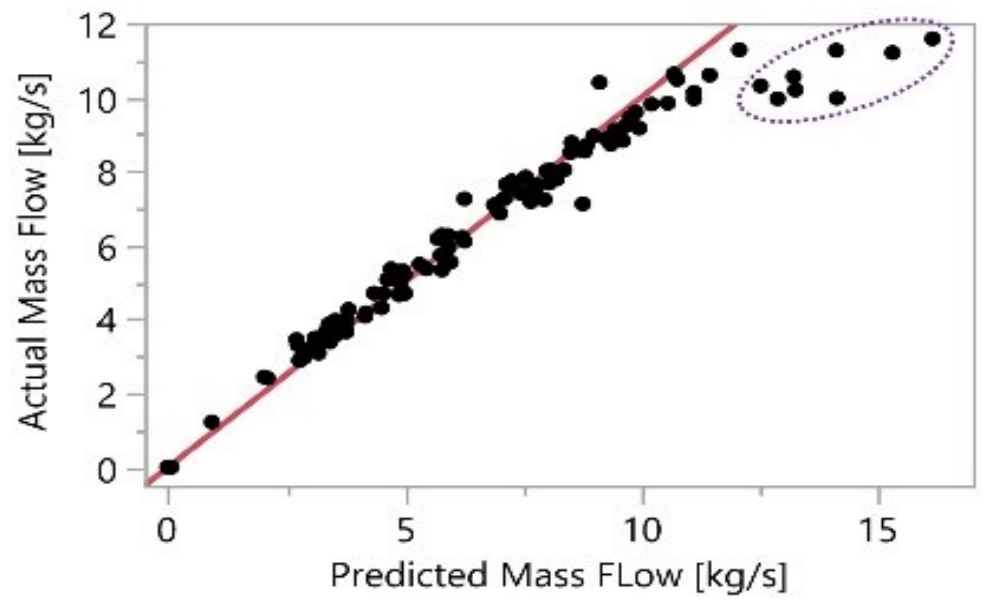

(a) Predicted Vs. Actual Mass Flow

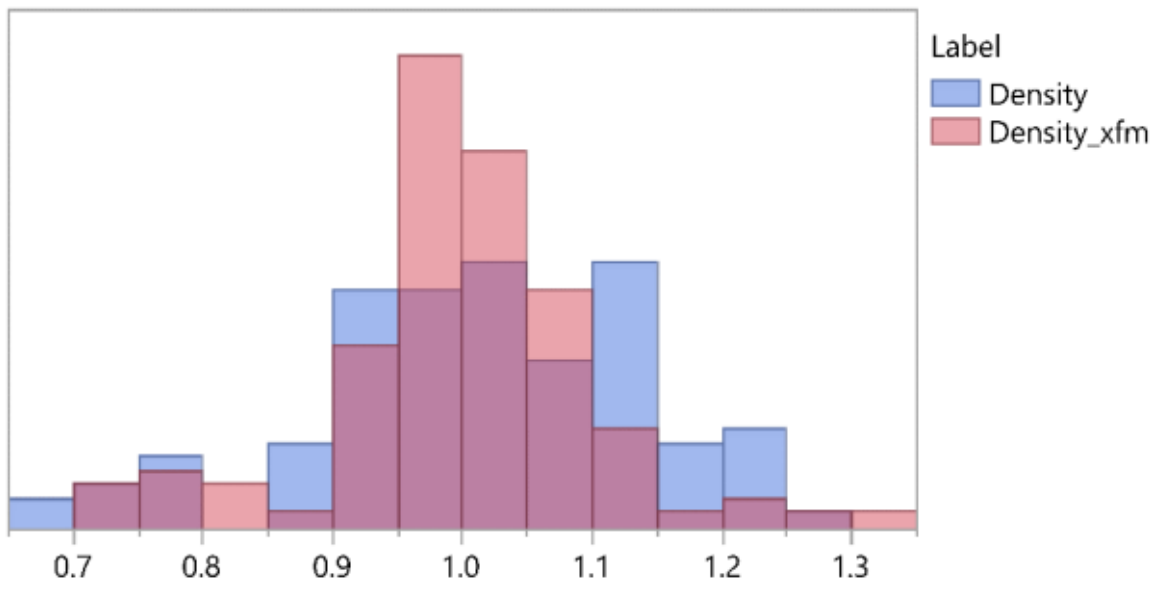

(b) Density Histograms

**Fig. 4** Predicted mass vs. actual mass flow shows a strong trend indicating the potential of this method as a yield monitor. The circled points represent flow levels beyond system capacity, where the bamboo begins to overflow the slats and slide back.

As predicting mass flow is an essential task of a yield monitor, the controlled lab testing provided a means to assess the sensor accuracy in this regard without introducing variation from external factors like material density changes. By utilizing a transformation on the volume to estimate density, the average predicted mass flow was plotted against the average actual mass flow for each run to observe correlation in Figure 4.a. With the intercept forced through zero, the linear correlation can be seen to be very strong, achieving an $R^2$ of 97.4% when runs with overflowing material are ignored. With these results showing strong potential, the system was considered good enough to pass to field testing.

### 3.2 Field Data Performance

The field data was observed to have changing density with volume flow trends similar to bamboo, although not the exact same trend, as seen in Figure 5.a. A subset of the data categorized by the season, crop type, and location is shown to emphasize these trends. Without coloring by these categories, such trends would not be obvious because unlike bamboo the harvested sugarcane is composed of a heterogeneous mixture of billets and trash, which causes the entire trends to translate vertically while still keeping their characteristic shape. Additionally, the particle density of the materials in the mixture changes based on factors like moisture and growing conditions for the season. Also, stalk thickness was observed to vary significantly even within a field, and the drum chopper which cuts the stalks into billets can have different numbers of blades that can impact billet length.

The aforementioned factors also interact with operation settings such as fan speed, which affects the mixture composition when trash is present. The fact that the density trends are visible with the different (location, season, and crop type) categories shows that density remains relatively stable for the same machine or group of similar machines operating in the same crop, region, and time period. To adjust for bulk density changes with volume flow, square root transformation (Equation 3) was applied to the volume to account for the packing effect of the material. The square root transform was applied to the measured volume prior to scaling by the accumulation term (the distance elevator slats moved between each measurement) to get the incremental volume. Figure 5.b shows the same data after applying a square root transform, and it can be seen that the trends have mostly disappeared within categories since they have been accounted for. Note that applying square root transform to volume translates into squashing or reducing the volume values, hence resulting in larger density values.

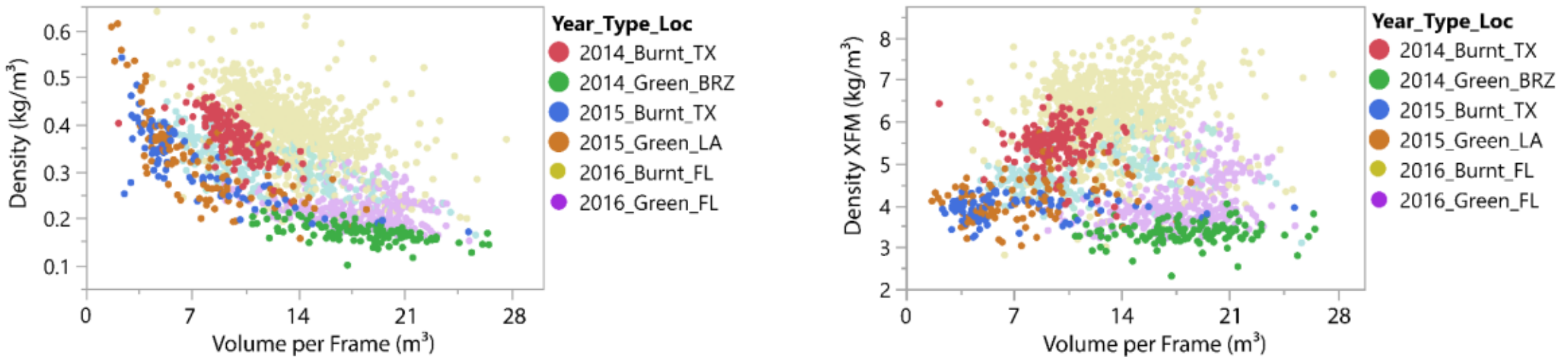

(a) Density Vs. Volume

(b) Density (with sqrt xfm) Vs. Volume

**Fig. 5** Before transformation of volume measured, clear trends can be seen of density decreasing with volume flows. **Right**: After transformation of the volume, trends have mostly disappeared within categories.

The large improvement in modeling out the trends can be seen quantitatively in the performance comparison between the CV column (w/o sqrt transform) and the "CV xfm" column (with sqrt transform) in Table 2.

**Table 2** Performance before and after applying square root transformation to the volume is given in terms of the coefficient of variation of the density values.

| Year | Type | Location | # Loads | CV (%) | CV (%) xfm |
|---|---|---|---|---|---|
| **2014** | **Burnt** | **TX** | 66 | 23.9 | 6.9 |
| **2014** | **Green** | **BRZ** | 166 | 28.2 | 12.7 |
| **2015** | **Burnt** | **TX** | 79 | 11.8 | 11.0 |
| **2015*** | **Green** | **LA** | 669 | **14.7** | **15.5** |
| **2016*** | **Burnt** | **FL** | 323 | 20.8 | 16.2 |
| **2016*** | **Green** | **FL** | 264 | 16.6 | 8.9 |

LA: Louisiana, TX: Texas, BRZ: Brazil, FL: Florida.
Tests conducted with very little system maintenance and across multiple machines*

Only in the case of 2015 green cane from LA was no improvement observed, but upon closer inspection of the time series plot of densities it was found that a shift in the density values occurred part way through the season, and the transform offered benefits if the different conditions are considered separately. During data collection it was observed that the burnt cane composition (and therefore density) could vary considerably within and between fields simply due to the levels of trash that are present after burning. Likewise, wet, leafy green cane tended to have lower density than dry green cane since the primary extractor fan could remove dry trash better than wet trash. Thus, it is concluded that a mechanism to quantify the levels of trash in the region of interest would be expected to provide a non-trivial enhancement in system accuracy across conditions.

From Table 2, transforming the volume to account for the stacking effect produced significantly better results in all cases except 2015 green cane harvested in Louisiana. Upon further investigation it was found that the density changed around a time when material consistency degraded due to operations and environment, as well as having more dirt and root balls showing up in point clouds. Hypothetically, covering the material in dirt could easily increase density and explain the fluctuations and overall net increase in density experienced there. The first month and a half of the season was very stable as can be seen prior to the vertical line indicating condition changes in Figure 6.a, which is a relatively long time. Thus, simply re-calibrating the density after the shift by tracking the volume of material that went into a semi-tractor trailer and obtaining a weight back from the mill could improve performance.

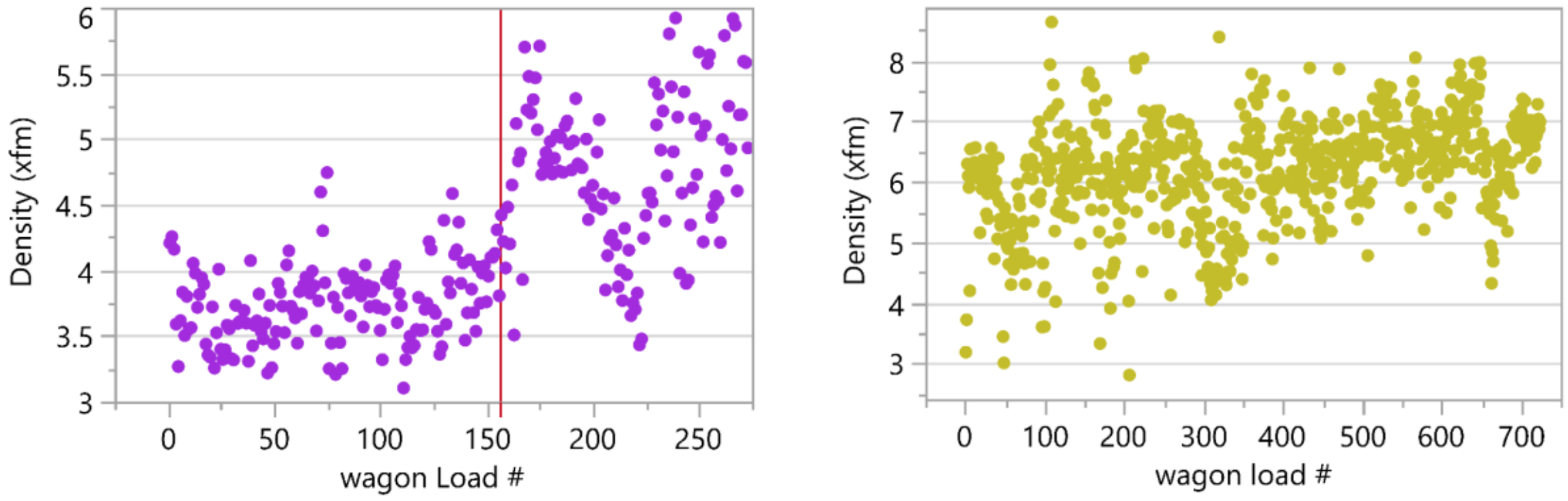

(a) Density over season in LA green cane 2015

(b) Density over season in FL burnt cane 2016

**Fig. 6** Left: After a month and a half of harvesting in the same region (LA), conditions changed and resulted in changes in the material density. The change is marked by a vertical line in the plot. The strong auto-correlation suggests shifting environmental and material conditions. Right: Large changes in trash levels within loads during 2016 FL harvesting caused overall higher variation within and between fields than other testing seasons.

Likewise with 2016 FL testing in burnt cane, a large variation in the amount of trash removed by burning was observed. This resulted in highly varying density over short periods of time that is separate from the stacking effect shown in Figure 6.b. Since the factors of dirt and leaves are both visually observable, it is projected that a mechanism to quantify them could be used to enhance density predictions by providing an adjustment based on the levels of these factors present. This is left for future work.

## 4 Conclusion

This work described a robust mass flow sensor for sugarcane harvesters using a stereo camera to estimate the volumetric surface of material on the elevator. The concept proved useful for estimating load weights, and could be used to characterize machine productivity as well. The concept is contingent on a calibrated value [density] to convert from volume to mass. The calibration values are relatively easy to obtain by tracking volume into a semi-tractor trailer and getting a weight back from the mill. When separated out by season, region, and crop type the accuracy is quite good, reaching CVs < 7% in some cases. Given the large density changes observed with varying trash levels, use of images to quantify and adjust for trash could have a very large impact on system performance and is the next iteration in line for the system. Thus by adjusting for trash levels, reaching a CV of <10% for the density values for all scenarios to match the more ideal performance of bamboo in the lab is deemed within reach for this concept and would maximize the value for estimating load weights, monitoring machine efficiency, and analyzing field productivity.